# A disorder-sensitive emergent vortex phase identified in high-$T_c$ superconductor (Li,Fe)OHFeSe

Dong Li[1,2,†,*], Peipei Shen[1,2,†], Jinpeng Tian[1,2], Ge He[1,2], Shunli Ni[1], Zhaosheng Wang[3], Chuanying Xi[3], Li Pi[3], Hua Zhang[1,2,7], Jie Yuan[1,7,8], Kui Jin[1,2,7,8], Evgeny F. Talantsev[5,6], Li Yu[1,2,7], Fang Zhou[1,2,7], Jens Hänisch[4], Xiaoli Dong[1,2,7,8*], and Zhongxian Zhao[1,2,7,8]

[1] *Beijing National Laboratory for Condensed Matter Physics, Institute of Physics, Chinese Academy of Sciences, Beijing 100190, People's Republic of China.*

[2] *School of Physical Sciences, University of Chinese Academy of Sciences, Beijing 100049, People's Republic of China.*

[3] *Anhui Province Key Laboratory of Condensed Matter Physics at Extreme Conditions, High Magnetic Field Laboratory of the Chinese Academy of Sciences, Hefei 230031, Anhui, People's Republic of China.*

[4] *Karlsruhe Institute of Technology, Institute for Technical Physics, Hermann-von-Helmholtz-Platz 1, 76344, Eggenstein-Leopoldshafen, Germany*

[5] *M.N. Mikheev Institute of Metal Physics, Ural Branch, Russian Academy of Sciences, 18 S Kovalevskoy St, Ekaterinburg 620108 Russia*

[6] *Nanotech Centre, Ural Federal University, 19 Mira St., Ekaterinburg 620002 Russia*

[7] *Songshan Lake Materials Laboratory, Dongguan, Guangdong 523808, People's Republic of China.*

[8] *Key Laboratory for Vacuum Physics, University of Chinese Academy of Sciences, Beijing 100049, People's Republic of China.*

[†] These authors contributed equally to this work.
[*] Correspondence to: lidong@iphy.ac.cn and dong@iphy.ac.cn;

## Abstract

The magneto-transport properties are systematically measured under *c*-direction fields up to 33 T for a series of single-crystal films of intercalated iron-selenide superconductor (Li,Fe)OHFeSe. The film samples with varying degree of disorder are grown hydrothermally. We observe a magnetic-field-enhanced shoulder-like feature in the mixed state of the high-$T_c$ (Li,Fe)OHFeSe films with weak disorder, while the feature fades away in the films with enhanced disorder. The irreversibility field is significantly suppressed to lower temperatures with the appearance of the shoulder feature. Based on the experiment and model analysis, we establish a new vortex phase diagram for the weakly disordered high-$T_c$ (Li,Fe)OHFeSe, which features an emergent dissipative vortex phase intermediate between the common vortex glass and liquid phases. The reason for the emergence of this intermediate vortex state is further discussed based on related experiments and models.

Disorder can significantly affect phase transitions in various systems. At the zero-temperature quantum critical point for example, disorder can induce the superconductor-insulator transition in quasi two-dimensional (2D) systems [1]. Besides such quantum phase transitions, disorder also has vital effects on classic thermodynamic phase transitions, especially in the vortex system arising from the mixed state of type-II superconductors. Although the dynamical behavior of vortices is mainly controlled by thermal fluctuations in high-$T_c$ superconductors (HTS's) owing to an intrinsically high Ginzburg-Levanyuk number ($Gi$), the statistical mechanics may be prominently influenced by the quenched disorder in real superconducting materials [2]. Generally, disorder transforms the ideal vortex lattice into a vortex glass (VG) state [3], which prevents the superconducting currents from causing dissipation, especially in applied magnetic fields. As a result, the first-order melting-transition $T_m(H)$ between vortex lattice and vortex liquid (VL) is replaced by the second-order glass-liquid-transition $T_g(H)$ between VG and VL, which is usually located nearby the original melting line [4,5]. The VG state was confirmed early on in experiment [6,7] and clarified in theory [3,8,9]. With increasing temperature, the VG phase continuously melts into the VL phase accompanied by the emergence of a finite resistance already at infinitesimally small currents, as commonly seen in HTS's [10,11].

On the other hand, previous theoretical [12] and experimental [13] studies have also proposed an additional vortex state that appears between the common VG and VL states. Such a vortex state, referred to as vortex slush state in ref. [13], is characteristic of losing the long-range correlations but keeping the short-range correlations among vortices, as well as a concomitant shoulder-like feature in resistance [13]. The vortex slush phase has mainly been reported for high-$T_c$ layered cuprate YBa$_2$Cu$_3$O$_7$ [4,13-17], but also for a low-$T_c$ organic superconductor [18]. Experimental [15,17] as well as theoretical work [19] has shown that this vortex phase is quite sensitive to disorder. So far, however, experimental information of the possible presence of intervening vortex phases in other high-$T_c$ superconducting systems is still inadequate, presumably due to the lack of high-quality samples and/or the difficulty in controlling their degree of disorder. The high-$T_c$ iron-selenide (Li,Fe)OHFeSe [20], formed by intercalating (Li$_{1-x}$Fe$_x$)OH molecules between adjacent superconducting FeSe layers, provides an excellent platform to investigate such an exotic vortex state, owing to the high tunability of its physical/chemical properties and the advance in its sample growth techniques [21-24].

In this paper, we report the systematic magneto-transport results obtained under *c*-direction fields up to 33 T on a series of superconducting (Li,Fe)OHFeSe films grown hydrothermally. Varying degree of disorder can be achieved for the hydrothermal films by appropriately adjusting their synthesis conditions. The level of disorder of each sample is characterized by the individual residual resistance ratio (RRR). Interestingly, in the high-$T_c$ (Li,Fe)OHFeSe samples with weak disorder, a magnetic-field-enhanced shoulder-like feature is clearly observed in the temperature-dependent resistance in the mixed state. The shoulder feature appears at a characteristic temperature $T'$ that is far above the critical glass region of the glass-liquid transition. With increasing level of disorder, the shoulder feature gradually fades away in the (Li,Fe)OHFeSe samples, demonstrating its fragility to disorder. Based on the experiment and model analysis, we establish a new vortex phase diagram for the weakly disordered high-$T_c$ (Li,Fe)OHFeSe showing the shoulder feature. This new phase diagram is distinguished from the commonly observed one by the emergence of an exotic dissipative vortex phase between the VG and VL phases. The reason for this emergent intermediate vortex state identified here is discussed based on related experiments and previously proposed models.

The (Li,Fe)OHFeSe single-crystal films were grown on $LaAlO_3$ substrates via matrix-assisted hydrothermal epitaxy (MHE) as reported elsewhere [22-24]. The level of disorder can be controlled by adjusting the synthesis temperature in the range from 120 to 180 °C [25]. The crystalline quality of all the hydrothermal (Li,Fe)OHFeSe films studied here is better than the bulk single crystals as well as iron-based films obtained by common deposition methods, as demonstrated by the results of x-ray diffraction [23] and other probes such as scanning tunneling microscopy (STM) [26-28] and electronic Raman spectroscopy [29]. The electrical transport properties were measured under magnetic fields along $c$-axis up to 16 T on a Quantum Design PPMS system with the standard four-probe method. The transport measurements under high $c$-axis magnetic fields up to 33 T were performed on the Steady High Magnetic Field Facilities. The superconducting transition $T_c$ of all the samples was defined as the temperature of 50% $R_n$ (the normal-state resistance just above superconducting transition). Some of the measuring parameters and properties of the studied samples are summarized in table 1.

Figure 1(a) shows the normalized temperature-dependent resistance near the superconducting transition ($T_c$ = 43.1 K) for sample S2. Despite having a high crystalline quality and/or weak disorder as indicated by its high RRR = 42, this sample exhibits a shoulder-like feature in the $R$-$T$ curves at lower temperatures in the mixed state. This feature is enhanced in the higher fields, while it becomes faint in the lower fields. We notice that, in the layered iron-based superconductor $CaKFe_4As_4$, intergrowth phases could significantly affect the vortex pinning properties [30]. In the present intercalate films of (Li,Fe)OHFeSe, however, we observe no indication of such planar defects by scanning transmission electron microscopy [31] or STM [26-28]. Moreover, as will be seen below, the shoulder feature is reproducibly observed in the high-quality (Li,Fe)OHFeSe films with high RRR and $T_c$ values, whereas it is undetectable in the films with much-reduced RRR and lower $T_c$ values. Therefore, the possibility of inhomogeneities or grain-boundary effects as the reasons for the shoulder feature can be ruled out for these high-$T_c$ samples.

For an understanding of the shoulder feature, we further analyze the resistance data in the mixed state based on the known vortex glass theory. The VG theory predicts that the resistivity vanishes at $T_g$ following the power law $\rho = \rho_0|T/T_g - 1|^s$ [6], where $\rho_0$ is a characteristic resistivity in the normal state and $s$ is a critical exponent. Therefore, in the critical region above the second-order transition from VG to VL phase, the reciprocal of the logarithmic resistance derivative with respect to temperature is linearly dependent on temperature as

$$\left(\frac{d\ln R}{dT}\right)^{-1} = \frac{1}{s}(T - T_g). \tag{1}$$

Figure 1(b) shows the linear region observed for sample S2 with weak disorder (RRR = 42), indicating the VG state below $T_g$. The dissipation in the linear fitting region is dictated by the ground state of the VG. Importantly, we find that a corresponding magnetic-field-enhanced hump shows up in $[d\ln R(T)/dT]^{-1}$ at a characteristic temperature $T'$ in the mixed state, and it vanishes with cooling just above the linear critical region. Here $T'$ is defined as the temperature at the hump peak (as indicated by the star symbol in figure 1(b)), which is the same temperature as the inflection point of $R(T)$ in figure 1(a) that is associated with the shoulder feature. The hump in $[d\ln R(T)/dT]^{-1}$ is readily discernible even at lower magnetic fields in the weakly disordered sample S2. In contrast, however, such a hump is absent at any field in sample S8 ($T_c$ = 32.5 K) with strong disorder (as indicated by its RRR = 3, which is one order of magnitude smaller than the RRR = 42 of sample S2), as can be seen from the corresponding resistance (figure 1(c)) and derivative (figure 1(d)) data. For sample S8, the resistance at lower $T$ is also well fitted by the VG theory, as shown in figure 1(d). Moreover, its monotonic increase in $[d\ln R(T)/dT]^{-1}$ at higher $T$ above the linear critical region is a typical feature of the VL state resulting from the thermally assisted flux motion, consistent with previous work [32]. In the case of sample S2 showing the shoulder in $R(T)$, however, this monotonic increase associated with the VL state is recovered only at a temperature slightly higher than $T'$, as can be seen from figure 1(b). Therefore, the appearance of the shoulder feature on cooling to $T'$ and its disappearance just above the linear critical region of VG state, combined with its magnetic-field-enhanced nature, imply an emergent exotic vortex phase between the VG and VL phases in the weakly disordered high-$T_c$ (Li,Fe)OHFeSe film samples. Further evidence for this emergent intermediate vortex phase will be discussed later.

In figure 2, we present the data of $[d\ln R(T)/dT]^{-1}$ as function of $T$ measured at 9 T for a series of (Li,Fe)OHFeSe films (samples S2 - S6). Their RRR is substantially reducing from 42 down to 7, while $T_c$ changes little between 43.2 and 42.1 K. It is evident that the hump above the linear critical region gradually fades away with decreasing RRR, and finally disappears at a much-reduced RRR value such as 7. For sample S4 with a moderate RRR = 23 (see figure 2), we have measured its *R-T* data under high magnetic fields along *c*-axis up to 33 T, shown in figure 3(a). Due to the severe noise of the high-field data, we have to plot in figure 3(b) the temperature-dependent derivative of resistance with respect to temperature to track the evolution of the shoulder feature in *R*(*T*). The characteristic $T'$ is located at the local minimum of the d*R*(*T*)/d*T* curves, as indicated by the star symbol in figure 3(b). One can see that the shoulder feature, which is smeared out at low fields, becomes pronounced at high fields and its characteristic $T'$ is concomitantly reduced (figure 3(b)). Thus, compared to the data at $\mu_0 H$ = 9 T (figure 2), the results at high fields up to 33 T further reveal that, in the high-$T_c$ sample S4 with the moderate RRR = 23, the shoulder feature only weakens rather than vanishes. Furthermore, we have also measured the high-field magneto-transport property by sweeping the *c*-direction field up to 33 T at various fixed temperatures, as shown in figures 3(c) and 3(d) for sample S1 ($T_c$ = 43.1 K) with high RRR = 50 and sample S7 ($T_c$ = 34.2 K) with much lower RRR = 5, respectively. The *R-H* data of the high-RRR S1 displays distinctive behavior in the shoulder temperature range (figure 3(c)), as compared to the counterpart of the low-RRR S7 without the shoulder feature (figure 3(d)). The upper critical field $H_{c2}(T)$ and irreversibility field $H_{irr}(T)$ are extracted at 50% $R_n$ and 0.1% $R_n$, respectively, as shown in figures 3(c) and 3(d).

By summarizing the results of $H_{c2}(T)$, $T'(H)$, $H_{irr}(T)$, and $T_g(H)$, we plot in figure 4(a) a new vortex phase diagram for the high-$T_c$ (Li,Fe)OHFeSe samples S1/S2 (50 ≥ RRR ≥ 42) and S4 (RRR = 23) with weak and moderate disorder, respectively. For comparison, we also present in figure 4(b) the commonly observed vortex phase diagram for sample S7 (RRR = 5) with strong disorder. Although $T_g(H)$, the true boundary of the VG state, is shown only for S2, $H_{irr}(T)$ can be used to estimate the border of the dissipationless VG state, which is slightly above the $T_g(H)$ line (see figure 4(a)). In the common phase diagram of figure 4(b), the dissipative VL phase exists in the region between $H_{irr}(T)$ and $H_{c2}(T)$; its nonzero resistance results from the thermally activated flux motion. In the new phase diagram of figure 4(a), in contrast, the exotic intermediate vortex phase emerges between the lines of $H_{irr}(T)$ and the characteristic temperature $T'$. Moreover, the emergence of this intermediate vortex state having finite resistance significantly suppresses the $H_{irr}(T)$ line to lower temperature/field, as compared to the $H_{irr}(T)$ line in figure 4(b) where the intermediate vortex state is absent. This observation is discussed in more detail as follows.

For a strongly layered HTS, *e.g.* cuprate $Bi_2Sr_2CaCu_2O_{8-\delta}$ [33,34], there will be a lower boundary of the vortex liquid phase, referred to as $T_m^{2D}$ if the magnetic field is significantly higher than a dimensional crossover field $H_{cr}$. For the strongly layered (Li,Fe)OHFeSe intercalate studied here (*e.g.* samples S1/S2 with $T_c \sim 43$ K), the maximum applied magnetic field of 33 T well exceeds a $\mu_0 H_{cr} \sim \phi_0/(\gamma d)^2 \sim 20$ T, estimated by using the values of the anisotropy $\gamma = 11$ [35] and the interlayer spacing $d = 9.32$ Å [22]. The high-field-limiting $T_m^{2D}$ is known to be proportional to $d/(\lambda_{ab}(0))^2$ [33], where $\lambda_{ab}$ is the in-plane superconducting penetration depth. According to our earlier study of (Li,Fe)OHFeSe films [23], $d = 9.29$ Å for sample S7 with $T_c = 34.2$ K. The ground-state in-plane penetration depth $\lambda_{ab}(0)$ was estimated as 160-200 nm for samples S1/S2 [35], and the ratio between the $(\lambda_{ab}(0))^{-2}$ values of samples S1/S2 (figure 4(a)) and sample S7 (figure 4(b)) can be roughly estimated from the ratio of their $T_c$ values (= 1.26), following a positive correlation between $T_c$ and $(\lambda_{ab}(0))^{-2}$ as reported in iron-based superconductors [36]. So, with the assumption of a high-field-limiting $T_m^{2D}$ for (Li,Fe)OHFeSe, the values of $T_m^{2D}$ for S1/S2 are to be larger than that of S7 by a factor of around 1.3, or at least they are not smaller than that of S7.

In the common vortex phase diagram of figure 4(b), we show the dashed vertical line at $T = 10$ K as the estimated asymptote of the $H_{irr}(T)$ line at high fields. The temperature scale of 10 K is expected to be close to the high-field-limiting $T_m^{2D}$ for sample S7. Correspondingly, the $T_m^{2D}$ value for samples S1/S2 in figure 4(a) can be roughly estimated as about 13 K. If no additional dissipative vortex state intervenes (like the case of figure 4(b)), the dissipationless VG state would have appeared just below $T_m^{2D} \sim 13$ K. In fact, however, the irreversibility line $H_{irr}(T)$ at high fields $\gtrsim 33$ T is suppressed to $T \lesssim 5$ K, well below the otherwise expected $T_m^{2D} \sim 13$ K for samples S1/S2, due to the emergent dissipative vortex phase, as shown in figure 4(a). Such a suppression of $H_{irr}(T)$ can be understood in terms of the emergent vortex state having lost the long-range correlations among vortices [13], which is distinct from the dissipationless VG state below the $H_{irr}(T)$ line.

Moreover, the observed suppression of the $H_{irr}(T)$ line is also consistent with the recent STM result of our high-$T_c$ (43 K) (Li,Fe)OHFeSe films. The STM experiment performed in the VG state at $\mu_0 H = 10$ T and $T = 4.2$ K [26] has clearly observed a finite number of free vortices on the FeSe layers. This demonstrates that the density of vortices exceeds the density of defects in the high-quality (Li,Fe)OHFeSe films (corresponding to the present S1 and S2 in figure 4(a)). Therefore, the STM observation gives a strong hint that the formation of the intermediate vortex phase is related mainly to the free vortices present in the high-$T_c$ (Li,Fe)OHFeSe samples with weak disorder (such as those presented in figure 4(a)). In view of this, we argue that, in the samples like S7 in figure 4(b) with enhanced disorder, the defect density is increased probably to such a level (*e.g.* close to or even exceeding the vortex density at any reasonable field) that the exotic dissipative vortex state is not able to form due to the absence of free vortices.

Here we emphasize that the formation of the intermediate vortex phase is not a phenomenon occurring within the VL region, as inferred from the above discussion of $T_\mathrm{m}^\mathrm{2D}$. Rather, it is governed mainly by the level of disorder, and this exotic vortex phase becomes pronounced as disorder is substantially reduced, as summarized in figure 4. The emergent vortex state above the lower limiting $H_\mathrm{irr}(T)$ line (figure 4(a)) may keep the short-range correlations among vortices, in accordance with the previous analysis in YBa$_2$Cu$_3$O$_{7-\delta}$ [13,17]. The short-range correlations among vortices finally vanish as crossing the upper limiting $T'$ line into the VL region. Further, we find that, compared to the case of samples S1/S2 with very weak disorder (50 ≥ RRR ≥ 42), the lower $H_\mathrm{irr}(T)$ line of sample S4 with moderate disorder (RRR = 23) is shifted toward the upper $T'$ line, as seen in figure 4(a). However, the $T'$ line itself seems insensitive to the change of disorder between the samples S1/S2 and S4 having similar $T_\mathrm{c}$ values, suggestive of the $T'$ similarity to the original $T_\mathrm{m}$ line that is robust against disorder [13]. Therefore, it appears that the lower $H_\mathrm{irr}(T)$ line eventually overlaps the upper $T'$ line at strong disorder, resulting in the commonly observed vortex phase diagram shown in figure 4(b).

Now we switch to discussing the exotic vortex phase in figure 4(a) in connection with previously proposed models, including the vortex slush phase as first discussed based on experiments on YBa$_2$Cu$_3$O$_{7-\delta}$ by Worthington *et al* [13]. The presently identified exotic vortex state resembles the previously proposed vortex slush state in terms of their intermediate state nature between the VG and VL states; namely, both of them lose the long-range correlations but keep the short-range correlations among vortices. However, significant differences exist between them as well. As already demonstrated, the exotic vortex state in the iron-selenide (Li,Fe)OHFeSe becomes more pronounced in the limit of weak disorder. This characteristic differs from the vortex slush state that survives only within a moderately disordered region in the defect-enhanced cuprate YBa$_2$Cu$_3$O$_{7-\delta}$ [13]. Moreover, the resistivity of the vortex slush state is about one order of magnitude smaller than the corresponding vortex liquid state [13]. This is also distinctly different from the present observation. Additionally, we note that a dimensional transition may also induce a dissipative vortex state below $T_\mathrm{m}^\mathrm{2D}$, as proposed for strongly layered HTS [2]. If this could apply to the intercalated HTS (Li,Fe)OHFeSe, the exotic vortex phase identified here would be a quasi-2D superfluid solid phase with pancake vortices [2,33,37]. However, this picture still lacks an account of why the intermediate dissipative vortex phase present in S1/S2 (figure 4(a)) is absent in S7 (figure 4(b)) that is also quasi-2D in nature.

Therefore, further studies are required to elucidate the detailed mechanism for the intermediate vortex phases. Especially in materials having weak disorder, the ground state should be Bragg glass [38,39] at low fields, which is transformed into the VG phase at higher fields [11,40]. The new vortex phase diagram in figure 4(a) does not delineate the Bragg glass at low fields. A further investigation into the low-field vortex phase diagram is currently underway. On the other hand, the magnetic-field-enhanced shoulder feature in the mixed state was previously detected not only in the high-$T_c$ intercalated iron-selenide (Li,Fe)OHFeSe [35], but also in other high-$T_c$ iron-selenide intercalates such as Li$_x$(NH$_3$)$_y$Fe$_2$Se$_2$ [41], (TBA)$_{0.3}$FeSe [42] and (EMIM)$_x$FeSe [43], as well as very recently in layered iron-arsenide superconductor RbCa$_2$Fe$_4$As$_4$F$_2$ ($T_c$ = 31 K) [44]. Moreover, a similar magnetic-field-broadened feature was reported before for high-$T_c$ layered cuprates YBa$_2$Cu$_3$O$_{7-\delta}$ (untwinned) [45] and Bi$_2$Sr$_2$CaCu$_2$O$_{8-\delta}$ [46]. Therefore, whether a similar dissipative intermediate vortex phase also exists in these high-$T_c$ superconductors is an important issue worthy of further investigation as well. In particular, the region of dissipationless VG phase is significantly suppressed due to the emergence of the dissipative intermediate vortex state. This is detrimental to potential high-field applications of the high-$T_c$ superconductivity. Fortunately, we have checked that the exotic intermediate vortex phase can be eliminated by introducing the transition metal Mn into the crystal structure, as done in our previous studies [31,47]: Doping Mn into (Li,Fe)OHFeSe films yields the common vortex phase diagram shown in figure 4(b). Meanwhile, the critical current density is strongly enhanced in the Mn-doped (Li,Fe)OHFeSe films compared to the undoped ones, and $T_c$ can also be kept at a high value of 36.6 K [31].

To conclude, we identify a dissipative vortex phase that emerges between the vortex glass and liquid phases in the films of high-$T_c$ iron-selenide superconductor (Li,Fe)OHFeSe, and establish a new vortex phase diagram. We show that this exotic vortex phase is sensitive to the level of disorder in (Li,Fe)OHFeSe, it becomes more pronounced in the high-$T_c$ films in the limit of weak disorder, while undetectable in the films with strong disorder. The emergence of the dissipative vortex phase corresponds to the presence of a significant number of free vortices in the weakly disordered samples. Nevertheless, it calls for further investigations to comprehensively understand the exotic intermediate vortex state, including its possible link with the quasi-two dimensionality, as well as the open issue of whether a similar intermediate vortex state ubiquitously exists in other high-$T_c$ layered superconducting compounds.

**Acknowledgements**

D. Li thanks Dr. W. Hu, professors Z. X. Shi, and H. H. Wen for helpful discussions. This work was supported by National Key Research and Development Program of China (Grant No. 2017YFA0303003), the National Natural Science Foundation of China (Grant Nos. 12061131005, 11834016, 11888101, and 11874359), the Strategic Priority Research Program of Chinese Academy of Sciences (XDB33010200 and XDB25000000). J. Hänisch thanks for the financial support provided by the Deutsche Forschungsgemeinschaft (DFG) through project HA6407/4-1. E. F. Talantsev thanks for the financial support provided by the Ministry of Science and Higher Education of Russia (theme "Pressure" No. AAAA-A18-118020190104-3) and by Act 211 Government of the Russian Federation, Contract No. 02.A03.21.0006. A portion of this work was performed on the Steady High Magnetic Field Facilities, High Magnetic Field Laboratory, Chinese Academy of Sciences, and supported by the High Magnetic Field Laboratory of Anhui Province.

**Reference**

[1] Goldman A. M. 2012 *Int. J. Mod. Phys. B* **24** 4081.
[2] Blatter G., Feigel'man M. V., Geshkenbein V. B., Larkin A. I., and Vinokur V. M. 1994 *Rev. Mod. Phys.* **66** 1125.
[3] Fisher M. P. 1989 *Phys. Rev. Lett.* **62** 1415.
[4] Safar H., Gammel P. L., Huse D. A., Bishop D. J., Lee W. C., Giapintzakis J., and Ginsberg D. M. 1993 *Phys. Rev. Lett.* **70** 3800.
[5] Nishizaki T. and Kobayashi N. 2000 *Supercond. Sci. Technol.* **13** 1.
[6] Koch R. H., Foglietti V. V., Gallagher W. J., Koren G., Gupta A., and Fisher M. P. 1989 *Phys. Rev. Lett.* **63** 1511.
[7] Sandvold E. and Rossel C. 1992 *Physica C* **190** 309.
[8] Fisher D. S., Fisher M. P., and Huse D. A. 1991 *Phys Rev B* **43** 130.
[9] Huse D. A., Fisher M. P. A., and Fisher D. S. 1992 *Nature* **358** 553.
[10] Brandt E. H. 1995 *Rep. Prog. Phys.* **58** 1465.
[11] Kwok W. K., Welp U., Glatz A., Koshelev A. E., Kihlstrom K. J., and Crabtree G. W. 2016 *Rep Prog Phys* **79** 116501.
[12] Larkin A. I. and Ovchinnikov Y. N. 1979 *J. Low Temp. Phys.* **34** 409.
[13] Worthington T. K., Fisher M. P., Huse D. A., Toner J., Marwick A. D., Zabel T., Feild C. A., and Holtzberg F. 1992 *Phys Rev B* **46** 11854.
[14] Reyes A. P., Tang X. P., Bachman H. N., Halperin W. P., Martindale J. A., and Hammel P. C. 1997 *Phys. Rev. B* **55** 14737.
[15] Nishizaki T., Shibata K., Naito T., Maki M., and Kobayashi N. 1999 *J. Low Temp. Phys.* **117** 1375.
[16] Wen H. H., Li S. L., Chen G. H., and Ling X. S. 2001 *Phys. Rev. B* **64** 054507.
[17] Shibata K., Nishizaki T., Sasaki T., and Kobayashi N. 2002 *Phys. Rev. B* **66** 214518.

[18] Sasaki T., Fukuda T., Nishizaki T., Fujita T., Yoneyama N., Kobayashi N., and Biberacher W. 2002 *Phys. Rev. B* **66** 224513.
[19] Nonomura Y. and Hu X. 2001 *Phys. Rev. Lett.* **86** 5140.
[20] Lu X. F. et al. 2015 *Nat. Mater.* **14** 325.
[21] Dong X. et al. 2015 *Phys. Rev. B* **92** 064515.
[22] Huang Y. et al. 2017 *Chin. Phys. Lett.* **34** 077404.
[23] Huang Y. et al. 2017 *arXiv: 1711.02920*.
[24] Dong X., Zhou F., and Zhao Z. 2020 *Front. Phys.* **8** 586182.
[25] Sun H. et al. 2015 *Inorg. Chem.* **54** 1958.
[26] Liu Q. et al. 2018 *Phys. Rev. X* **8** 041056.
[27] Chen C., Liu Q., Zhang T. Z., Li D., Shen P. P., Dong X. L., Zhao Z.-X., Zhang T., and Feng D. L. 2019 *Chin. Phys. Lett.* **36** 057403.
[28] Zhang T. et al. 2021 *Phys. Rev. Lett.* **126** 127001.
[29] He G., Li D., Jost D., Baum A., Shen P. P., Dong X. L., Zhao Z. X., and Hackl R. 2020 *Phys. Rev. Lett.* **125** 217002.
[30] Ishida S. et al. 2019 *npj Quantum Materials* **4** 27.
[31] Li D. et al. 2019 *Supercond. Sci. Technol.* **32** 12LT01.
[32] Yi X., Wang C., Tang Q., Peng T., Qiu Y., Xu J., Sun H., Luo Y., and Yu B. 2016 *Supercond. Sci. Technol.* **29** 105015.
[33] Glazman L. I. and Koshelev A. E. 1991 *Phys Rev B* **43** 2835.
[34] Chen B., Halperin W. P., Guptasarma P., Hinks D. G., Mitrović V. F., Reyes A. P., and Kuhns P. L. 2007 *Nat. Phys.* **3** 239.
[35] Hänisch J., Huang Y., Li D., Yuan J., Jin K., Dong X., Talantsev E., Holzapfel B., and Zhao Z. 2020 *Supercond. Sci. Technol.* **33** 114009.
[36] Bhattacharyya A., Adroja D. T., Smidman M., and Anand V. K. 2018 *Sci. China-Phys. Mech. Astro.* **61** 124702.
[37] Clem J. R. 1991 *Phys Rev B* **43** 7837.
[38] Giamarchi T. and Le Doussal P. 1995 *Phys Rev B* **52** 1242.
[39] Klein T., Joumard I., Blanchard S., Marcus J., Cubitt R., Giamarchi T., and Le Doussal P. 2001 *Nature* **413** 404.
[40] Giamarchi T. and Le Doussal P. 1997 *Phys. Rev. B* **55** 6577.
[41] Sun S., Wang S., Yu R., and Lei H. 2017 *Phys. Rev. B* **96** 064512.
[42] Shi M. Z. et al. 2018 *New J. Phys.* **20** 123007.
[43] Wang J., Li Q., Xie W., Chen G., Zhu X., and Wen H.-H. 2021 *Chin. Phys. B* **30** 107402.
[44] Xing X., Yi X., Li M., Meng Y., Mu G., Ge J., and Shi Z. 2020 *Supercond. Sci. Technol.* **33** 114005.
[45] Kwok W. K., Fleshler S., Welp U., Vinokur V. M., Downey J., Crabtree G. W., and Miller M. M. 1992 *Phys. Rev. Lett.* **69** 3370.
[46] Palstra T. T. M., Batlogg B., van Dover R. B., Schneemeyer L. F., and Waszczak J. V. 1990 *Phys. Rev. B* **41** 6621.
[47] Li D. et al. 2021 *Chin. Phys. B* **30** 017402.

Table 1. $T_c$, RRR, bridge width, film thickness, and applied current density $J_{app}$ of the studied samples.

| Sample | $T_c$ (K) | RRR | Bridge width (μm) [a] | Film thickness (nm) [b] | $J_{app}$ (A cm$^{-2}$) [c] |
|---|---|---|---|---|---|
| S1 | 43.1 | 50 | 20 | ~330 | ~1500 |
| S2 | 43.2 | 42 | 20 | ~950 | ~520 |
| S3 | 43.1 | 33 | - | - | ~20 |
| S4 | 42.5 | 23 | - | - | ~20 |
| S5 | 42.4 | 16 | - | - | ~20 |
| S6 | 42.1 | 7 | - | - | ~20 |
| S7 | 34.2 | 5 | 20 | ~500 | ~1000 |
| S8 | 32.5 | 3 | 50 | ~500 | ~400 |

[a]. All samples were measured with the standard four-probe method. Part of the samples were patterned with a microbridge to determine $J_c$ [31].

[b]. The film thickness was characterized by scanning electron microscopy.

[c]. The $J_{app}$ values were calculated with a 100 μA current applied in all the cases. For samples S3-S6, the sample width and thickness were roughly estimated as 1 mm and 500 nm, respectively.

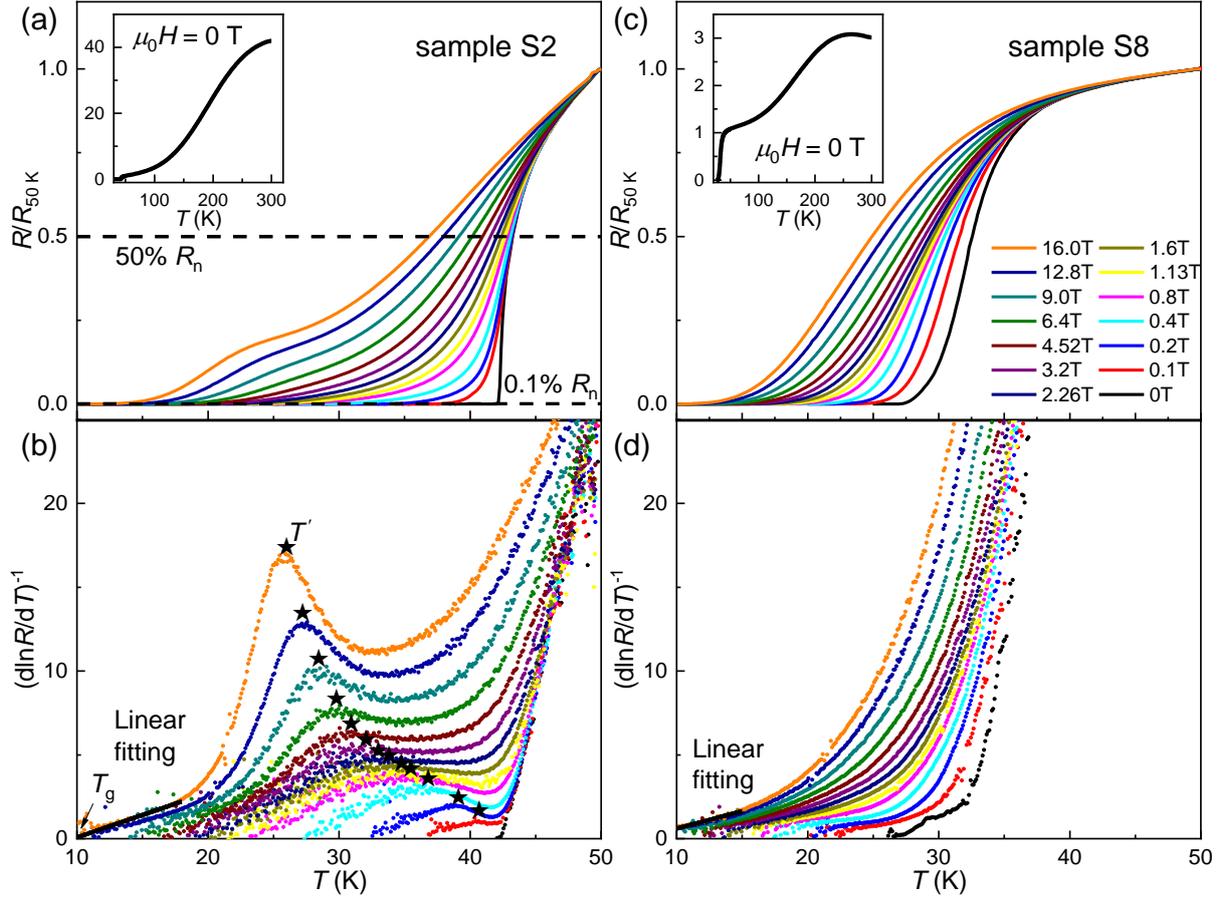

Figure 1. (a) and (b) show the normalized $R(T)/R_{50\,K}$ and $(d\ln R(T)/dT)^{-1}$ curves, respectively, for the high-$T_c$ (43.2 K) (Li,Fe)OHFeSe film sample S2 with RRR = $R_{300\,K}/R_{50\,K}$ = 42. The data are taken around the superconducting transition under various c-axis magnetic fields up to 16 T. A magnetic-field-enhanced shoulder feature is clearly observed. The black stars in (b) mark the characteristic temperature $T'$. (c) and (d) are the corresponding data for sample S8 (RRR = 3, $T_c$ = 32.5 K) without the shoulder feature. The zero-field $R(T)/R_{50\,K}$ curves up to 300 K are given in the insets.

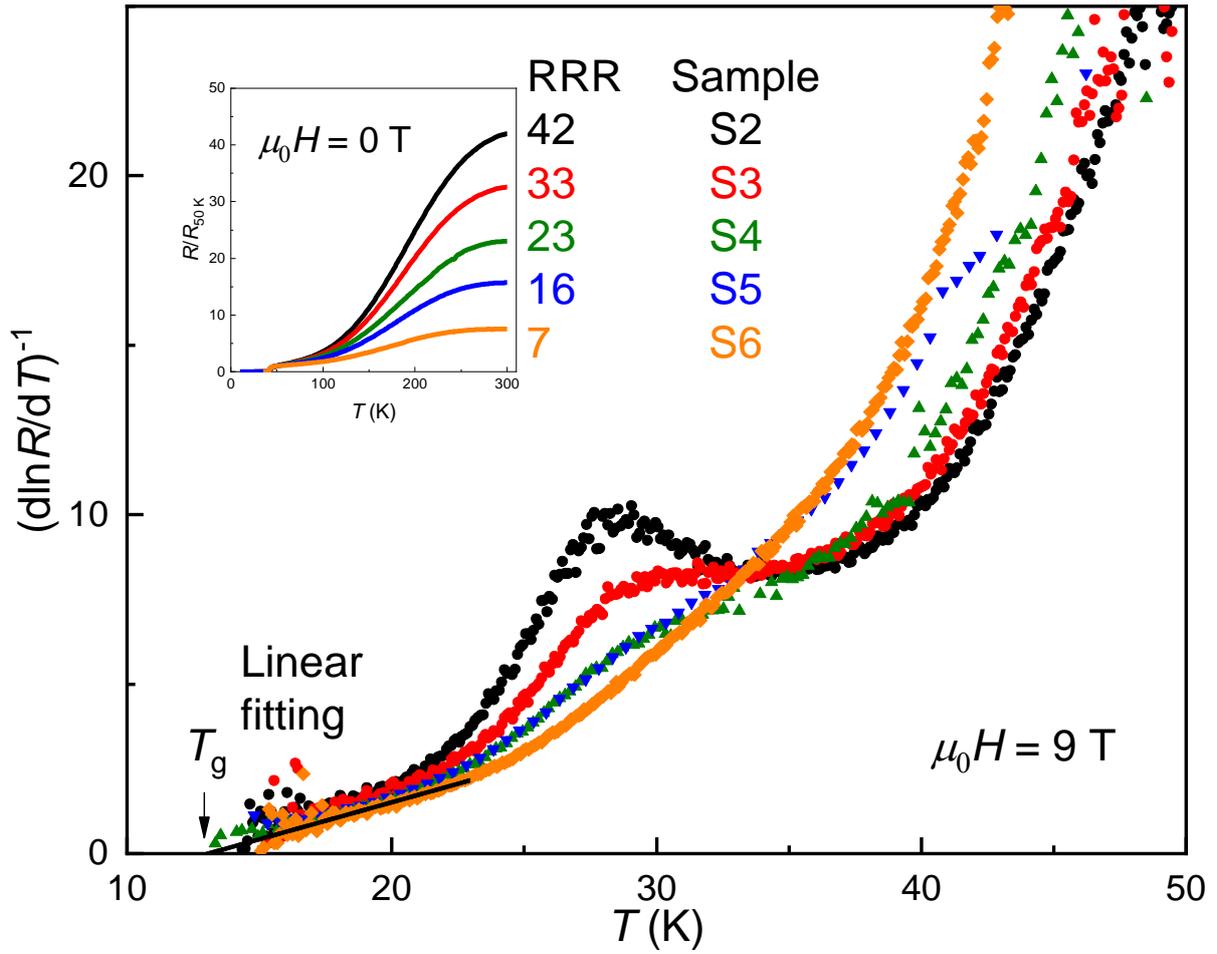

Figure 2. The temperature dependent $(d\ln R(T)/dT)^{-1}$ curves for a series of high-$T_c$ (43.2 – 42.1 K) (Li,Fe)OHFeSe films (S2 – S6) with different individual RRR = $R_{300\,K}/R_{50\,K}$ values (42 – 7). The data are measured under a fixed field of 9 T. The inset shows the zero-field $R(T)/R_{50\,K}$ curves up to $T$ = 300 K for the samples.

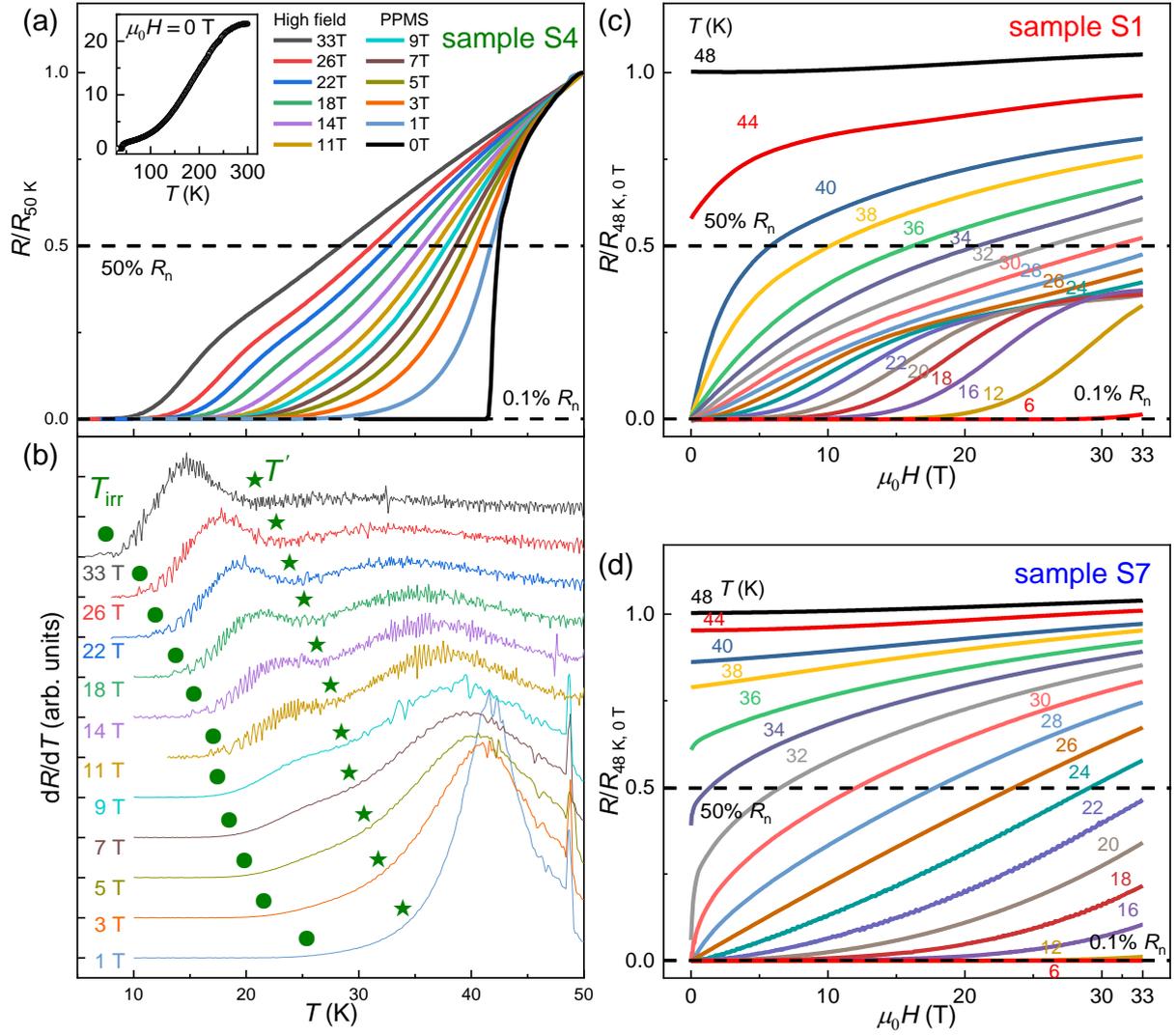

Figure 3. (a) The temperature dependence of normalized $R(T)/R_{50\,K}$ around the superconducting transition ($T_c$ = 42.5 K) for the (Li,Fe)OHFeSe film S4. The data are measured under various magnetic fields along $c$-axis up to 33 T. The inset shows the zero-field $R(T)/R_{50\,K}$ up to 300 K, with RRR = $R_{300\,K}/R_{50\,K}$ = 23. (b) The temperature-dependent resistance derivatives with respect to temperature for S4. The curves are offset for clarity. The olive stars and dots mark the positions of $T'$ and $T_{irr}$, respectively. (c) and (d) are the magnetic-field dependences of normalized resistance up to 33 T measured at various fixed temperatures for the films S1 (RRR = 50, $T_c$ = 43.1 K) and S7 (RRR = 5, $T_c$ = 34.2 K), respectively. The applied magnetic field is parallel to the $c$ axis. Here RRR = $R_{280\,K}/R_{48\,K}$ for S1 and S7.

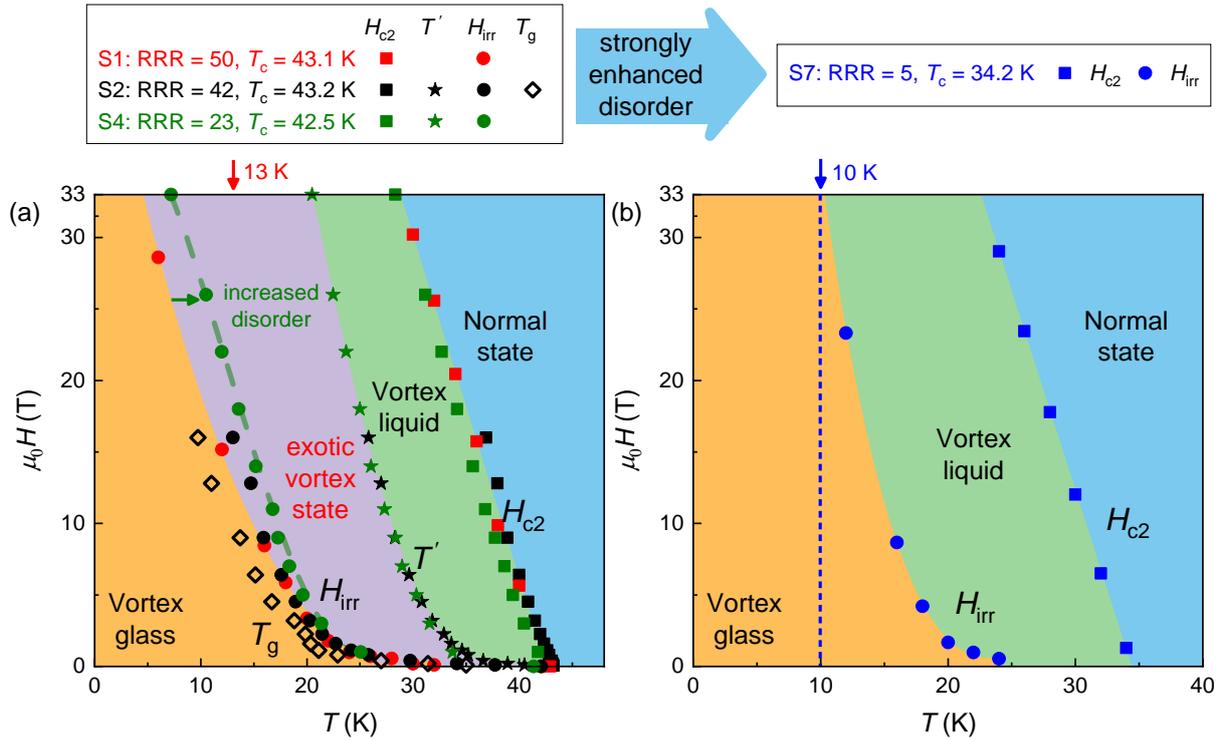

Figure 4. (a) The field-temperature vortex phase diagram for the (Li,Fe)OHFeSe films S1/S2 and S4 with very and relatively weak disorder, respectively. This phase diagram is characteristic of an exotic dissipative vortex phase that emerges between the vortex glass and liquid phases. (b) The commonly observed vortex phase diagram, i.e. without the emergent intermediate vortex phase, for the (Li,Fe)OHFeSe film S7 with strong disorder. The dashed vertical line at $T = 10$ K is the estimated asymptote of the $H_{irr}(T)$ line at high fields, i.e. with the temperature scale of 10 K expected to be close to the high-field-limiting $T_m^{2D}$. The lines of $H_{c2}(T)$, $T'(H)$, $H_{irr}(T)$, and $T_g(H)$ in the phase diagrams are represented with the solid squares, stars, dots, and hollow diamonds, respectively.